\documentclass[aps,prc,twocolumn,groupedaddress,showpacs]{revtex4}
\usepackage{graphicx,amsmath}

\newcommand{\jpsi}{\rm J/$\psi$}

\begin{document}

\def\torino{$^{1}$}
\def\alessandria{$^{2}$}


\title{Role of parton shadowing in the comparison of \mbox{p-A} and \mbox{A-A} results \\
\bf {on J/$\psi$ \bf suppression at energies available at the CERN Super Proton Synchrotron}}


\author{R.~Arnaldi\torino, 
P.~Cortese\torino\alessandria~and
E.~Scomparin\torino
}
\affiliation{
\torino \mbox{INFN, I-10125 Torino,~Italy}\\
\alessandria \mbox{Universit\`a del Piemonte Orientale, I-15121 Alessandria,~Italy}\\
}


\date{\today}

\begin{abstract}
The observation of an anomalous J/$\psi$ suppression in nucleus-nucleus collisions 
is one of the most important results of the SPS heavy-ion program. 
An essential ingredient in this result is the determination, obtained by 
studying \mbox{p-A} collisions, of effects not related with the formation of a
deconfined medium. These effects are extrapolated to \mbox{A-A} collisions,
determining a reference J/$\psi$ yield which is then compared with the measurements.
In this article we investigate the role of parton shadowing on the determination
of such a reference, and we calculate its effect for \mbox{In-In} and \mbox{Pb-Pb}
collisions as a function of rapidity and centrality. 
\end{abstract}


\maketitle


\section{Introduction}

The suppression of charmonium states was proposed a long time ago as a signature of the production of a deconfined state in
nucleus-nucleus collisions \cite{Sat86}. However, it was very soon realized that nuclear effects not related to deconfinement or, 
more generally, to the production of a hot medium may influence the observed charmonia suppression~\cite{Ger88,Kha97}. Such effects were investigated
through the study of charmonia production in \mbox{p-A} collisions. In these reactions, the produced $c\overline c$ pair may interact 
with the cold nuclear medium of the target nucleus, hindering the formation of a bound state.

Several \mbox{p-A} data samples exist today for \jpsi\ production at fixed target energies, in particular from NA50 at the 
SPS~\cite{Ale03,Ale04,Ale06}, from E866 at FNAL~\cite{Lei00} and from the HERA-B Collaboration at HERA~\cite{Abt09}.
Nuclear effects are usually 
parametrized by comparing the yields for various nuclear targets in a certain kinematical region, and then fitting their 
$A$ dependence in terms of the simple power law $A^\alpha$. Alternatively, the data are analyzed in the framework of the Glauber 
model~\cite{Gla59},
and their $A$ dependence expressed as an effective ``absorption'' cross-section $\sigma_{\rm J/\psi}^{eff}$. 
The interpretation of \jpsi\ results
in \mbox{p-A} collisions is the object of a rather strong theory effort. Nowadays, it is clear that not only the nuclear 
dissociation of the $c\overline c$ pair plays a role, but also effects like shadowing, initial- and final-state parton energy loss, and
possibly the intrinsic charm component of the projectile should be taken into account in a realistic description of the
results~\cite{Vog00}.

When studying \jpsi\ suppression in \mbox{A-A} collisions, a precise knowledge of nuclear effects is an essential requisite to 
disentangle genuine hot-medium effects. In the approach commonly used up to now~\cite{Kha97}, the effective quantity 
$\sigma_{\rm J/\psi}^{eff}$ is 
obtained analyzing \mbox{p-A} data taken in the same kinematic domain of \mbox{A-A} collisions under study. Then, it is assumed that 
in both \mbox{p-A} and \mbox{A-A} collisions nuclear effects, parametrized through the quantity $\sigma_{\rm J/\psi}^{eff}$, scale with $L$, 
the mean thickness of nuclear matter seen by the $c\overline c$ pair in its way through the projectile and target nuclei. 
In this way it is possible to determine the expected \jpsi\ yield for nuclear collisions as a function of centrality~\cite{Ale04}.
With this approach, a significant anomalous suppression (i.e., a suppression which goes beyond the estimated contribution from nuclear
effects) has been detected at SPS energies~\cite{Ale05,Arn07}.

In this article, we go a step further by taking explicitly into account parton shadowing in the determination of the 
cold nuclear matter effects in \mbox{A-A} collisions. Shadowing of partons in nuclei is a depletion of their population at
small momentum fraction of the nucleon, $x$, compared to that in a free nucleon, with a corresponding enhancement at moderate $x$
(anti-shadowing). Contrarily to final state dissociation of the $c\overline c$ pair, shadowing 
is not expected to scale with $L$, because in \mbox{p-A} collisions only partons in the target are affected by shadowing,
whereas in \mbox{A-A} the projectile is also involved. Therefore, a different approach to the evaluation of nuclear effects is required.

\section{Shadowing effects: \lowercase{p}-A collisions}

To give an estimate of shadowing effects on \jpsi\ production, we use the Color Evaporation Model (CEM) at leading order 
(LO)~\cite{Glu78}.
In this approach the charmonium production cross section for \mbox{p-A} collisions is obtained by integrating the free $c\overline c$ 
cross section from energy threshold to the open charm threshold. In absence of final-state interactions of the produced \jpsi\ one 
has~\cite{Vog99}

\begin{equation}
\frac{1}{A}\frac{d\sigma_{\rm J/\psi}}{dx_F}=2F\int_{2m_c}^{2m_D}mdm\frac{H_{pA}(x_1,x_2,m^2)}{\sqrt{s}\sqrt{x_F^2s+4m^2}}
\label{eq:1}
\end{equation}

where $F$ is the fraction of $c\overline c$ pairs which gives a \jpsi\ in the final state and $H_{pA}$ is given by

\begin{equation}
\begin{split}
H_{pA}(x_1,x_2,m^2)=f_g^p(x_1,m^2)f_g^A(x_2,m^2)\sigma_{gg}(m^2)+\\ +\sum_{q=u,d,s}f_q^p(x_1,m^2)f_{\overline q}^A(x_2,m^2)\sigma_{q\overline
q}(m^2)
\end{split}
\label{eq:2}
\end{equation}

$H_{pA}$ is the sum of two terms, corresponding to the elementary $c\overline{c}$ production by gluon fusion and  $q\overline{q}$
annihilation, convoluted with the parton densities $f_i^A(x,Q^2)$ and $f_i^p(x,Q^2)$ in the target and projectile nucleons, evaluated at $Q=m$. 
$x_1$ ($x_2$) is the fraction of the nucleon momentum carried by the parton of the projectile (target) which interacts.  
We have used in our calculation the GRV98LO~\cite{Glu98} set of parton distribution functions (PDF). The PDF for 
nucleons inside nuclei are modified with respect to the free ones according to the following expression:

\begin{equation}
f_i^A(x,Q^2)=R_i^A(x,Q^2)f_i^N(x,Q^2)
\label{eq:3}
\end{equation}

Various parametrizations of the PDF modifications from nuclear effects exist. We have used
the EKS98~\cite{Esk99} and EPS08~\cite{Esk08} sets, which are available for all mass numbers and have been implemented in the frame of 
the commonly used LHAPDF interface~\cite{Wha05}. 
The calculation has been performed for several A values, corresponding to Be, Al, Cu, In, W, Pb and U target nuclei.
For each target the ratio between the \jpsi\ production cross section (per nucleon-nucleon collision) in \mbox{p-A} and \mbox{p-p}, which we
refer to as shadowing factor in the following, is given by

\begin{equation}
S_{pA}^{\rm J/\psi}(x_F)=\frac{1}{A}\frac{d\sigma_{\rm J/\psi}^{pA}/dx_F}{d\sigma_{\rm
J/\psi}^{pp}/dx_F}
\label{eq:4}
\end{equation}

\begin{figure}[h]
\centering
\resizebox{0.48\textwidth}{!}
{\includegraphics*[bb=0 0 530 530]{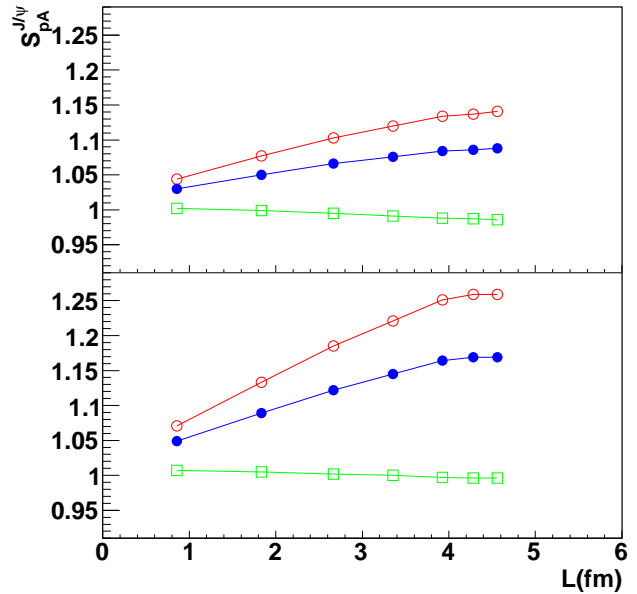}}
\caption{Shadowing factors for pA collisions at 158 GeV, at midrapidity. Open circles refer to the $gg$ fraction of the cross section,
open squares to $q\overline q$, solid circles to the total cross section. The top plot has been obtained with the EKS98 set, the bottom
plot with EPS08.}
\label{fig:1}
\end{figure}

In Fig.~\ref{fig:1} we present the result of the calculation of the shadowing factors for \mbox{p-A} collisions at 158 GeV, the energy 
used in the \mbox{A-A} data taking at the SPS. The plots refer to midrapidity, and we separately show the shadowing factors for 
the $gg$ and $q\overline q$ fraction of the production cross section. We plot the results as a function of the $L$ variable~\cite{Sha01}, 
computed for each nucleus using the Glauber model with realistic nuclear density distributions~\cite{DeV87}. 

\begin{figure}[h]
\centering
\resizebox{0.48\textwidth}{!}
{\includegraphics*[bb=0 0 530 530]{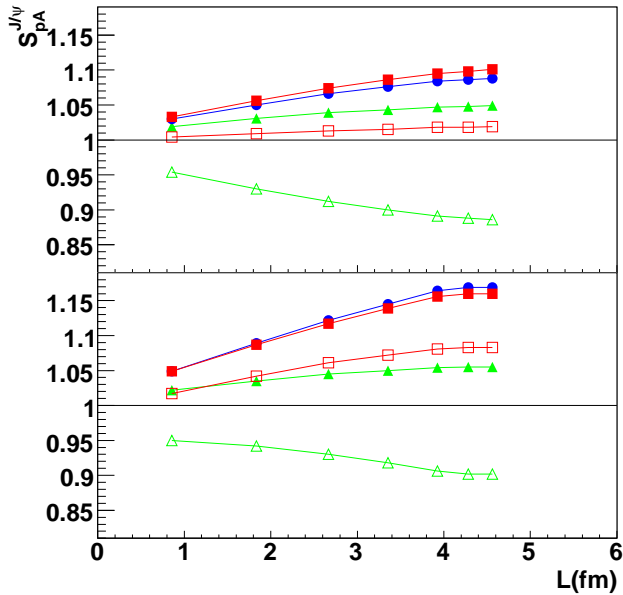}}
\caption{Shadowing factors for \jpsi\ production in \mbox{p-A} collisions at 158 GeV, using EKS98 (top plot) and EPS08 (bottom plot). 
Solid and open triangles refer to $y=1$ and  $y=-1$ respectively, solid and open squares to $y=0.5$ and $y=-0.5$, circles to $y=0$.}
\label{fig:2}
\end{figure}

\begin{figure}[h]
\centering
\resizebox{0.48\textwidth}{!}
{\includegraphics*[bb=0 0 530 530]{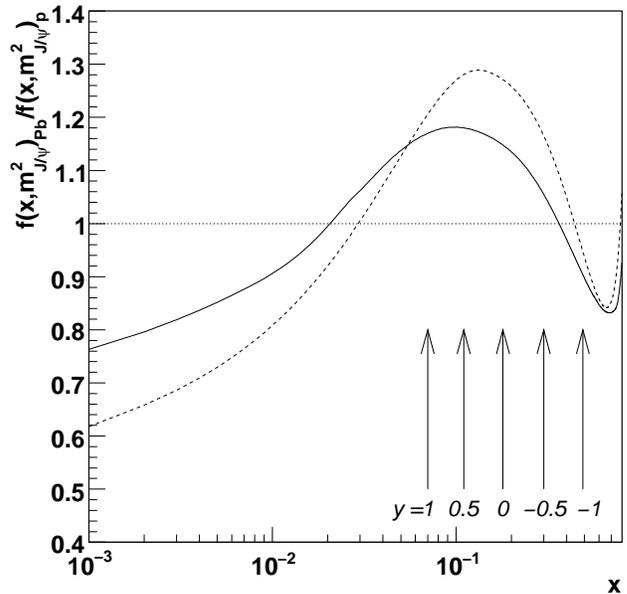}}
\caption{The EKS98 (continuous line) and EPS08 (dashed line) parametrizations of nuclear modifications to the PDFs for the Pb nucleus, 
calculated at $Q=m_{\rm J/\psi}$.
The arrows correspond to the rapidity values where the calculation of Fig.~\ref{fig:2} has been performed.}
\label{fig:2new}
\end{figure}

It can be noted that both EKS98 and EPS08 sets give shadowing factors larger than 1, and that such an anti-shadowing is more pronounced
when using EPS08. The $q\overline q$ fraction of the cross section exhibits a small shadowing, which is more than counterbalanced by the
dominant $gg$ production process. We have also verified, by choosing another
set of PDFs (MRST2001LO~\cite{Mar02}) that our results do not depend, within at most 2\%, on the specific choice of the PDF set. The flattening
of $S_{pA}^{\rm J/\psi}$ for heavy nuclei when using EPS08 is because of the fact that a constant shadowing is assumed beyond A=208~\cite{Sal09}. 
In Fig.~\ref{fig:2}  we present the results of a similar calculation at slightly forward ($y=0.5$, $y=1$) and backward ($y=-0.5$, $y=-1$)
rapidity.  We also plot the $y=0$ result for comparison. We recall that the region $0<y<1$ is the one where \jpsi\ results are 
available at SPS energies for \mbox{A-A} collisions. The shadowing factors are found to depend on $y$, as expected, 
since $x_2$ is directly related to this quantity. We recall that $x_2=(m_{\rm J/\psi}/\sqrt{s})\exp(-y)$ for \jpsi\ 
production in the CEM at LO. In Fig.~\ref{fig:2new} we plot the EKS98 and EPS08 parametrizations of nuclear effects on PDFs for the Pb
nucleus, relative to $Q=m_{\rm J/\psi}$. 
The arrows indicate the $x_2$ values corresponding to the rapidities where we have performed the shadowing factor
calculation of Fig.~\ref{fig:2}. It can be clearly seen that from $y=-1$ towards $y=1$ one goes from a shadowing to an anti-shadowing regime,
reaching a maximum of the anti-shadowing effect at $x_2\sim$0.10 for the EKS98 set ($\sim$0.13 for EPS08), which then decreases going towards
smaller $x_2$ (corresponding to more forward rapidities). Such an evolution can clearly be identified looking, at fixed $L$, to the shadowing
factors shown in Fig.~\ref{fig:2}, which are $<$1 at $y=-1$, then increase and finally decrease at forward rapidity.

We note that the shadowing factors calculated for $y=-0.5, -1$ are identical, for symmetry reasons, to those for \mbox{A-p} collisions 
at $y=0.5, 1$. Results on \mbox{A-p} collisions are interesting in order to study effects
from PDF modifications in the projectile, which will be important when studying \mbox{A-A} interactions. We see, for example, that at 
$y=1$ the shadowing effects in \mbox{A-p} are very different with respect to \mbox{p-A}. 


\section{Shadowing effects: \mbox{A-A} collisions}

When moving to \mbox{A-A} collisions, the cross section and the shadowing factors, integrated over centrality, can be calculated with 
an expression similar to Eq.~\ref{eq:1}, by replacing $H_{pA}(x_1,x_2,m^2)$ with $H_{AA}(x_1,x_2,m^2)$, taking now into account that 
shadowing affects both projectile and target nuclei. The problem becomes more complicate if shadowing factors have to be calculated for
various centrality intervals. Clearly, for various geometries of the collision, either the halo or the core of the nuclei will be mainly
involved, and the shadowing effects will be more important in the core than in the halo. Various parametrizations of the local shadowing
inside the nucleus have been proposed. We have used two of them~\cite{Kle03}:

\begin{equation}
R_{i,\rho}^A(x,Q^2,\vec{r},z)=1+N^A_\rho(R_i^A(x,Q^2)-1)\frac{\rho_A(\vec{r},z)}{\rho_0}
\label{eq:5}
\end{equation}

\noindent{and}
\begin{equation}
R_{i,L}^A(x,Q^2,\vec{r},z)=1+N^A_L(R_i^A(x,Q^2)-1)\frac{\int dz\rho_A(\vec{r},z)}{\int dz\rho_A(0,z)}
\label{eq:6}
\end{equation}

In the first one, shadowing in a certain location $(\vec{r},z)$ inside the nucleus is proportional to the local nuclear density 
$\rho_A(\vec{r},z)$, while in the second it is proportional to the length $L$ of nuclear matter crossed by the parton on its way 
through the nucleus. The normalization $N^A_\rho$ is fixed to ensure that 
$\int d\vec{r}dz R_{i,\rho}^A(x,Q^2,\vec{r},z)=R_i^A(x,Q^2)$ (and similarly for $N^A_L$).

\begin{figure}[h]
\centering
\resizebox{0.48\textwidth}{!}
{\includegraphics*[bb=0 0 530 530]{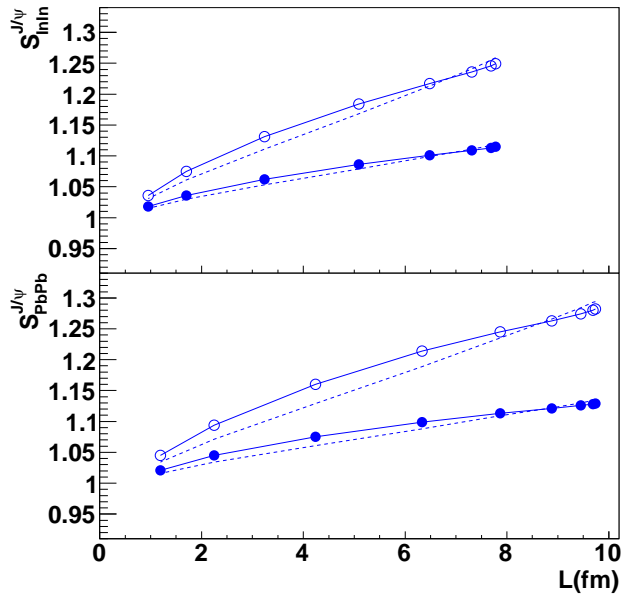}}
\caption{Shadowing factors for \mbox{In-In} (top) and \mbox{Pb-Pb} collisions at 158 GeV, at $y$=0.5, obtained with EKS98. Solid circles 
refer to the EKS98 parametrization, open circles to EPS08. The symbols connected by a
continuous line have been calculated using Eq.~\ref{eq:5}, whereas the dashed lines have
been obtained using the parametrization of Eq.~\ref{eq:6} for the local dependence of shadowing.}
\label{fig:3}
\end{figure}

The study of the shadowing factors in \mbox{A-A} collisions as a function of centrality has been performed for \mbox{In-In} and
\mbox{Pb-Pb} at 158 GeV/nucleon (see also~\cite{Eme99} for a previous investigation of the influence of shadowing on the centrality dependence 
of J/$\psi$ and Drell-Yan cross sections).  
A large number of events has been generated for centrality values in the interval 0$<b<$12 fm for
\mbox{In-In} and 0$<b<$16 fm for \mbox{Pb-Pb}, in steps of 2 fm, with a Glauber Monte-Carlo approach. We have used
$\sigma_{inel}^{pp}$=30 mb and the measured nuclear density distributions for In and Pb~\cite{DeV87}. For every \mbox{N-N} collision in each 
\mbox{A-A} interaction, we calculate $\rho_A(\vec{r},z)$ and $L(\vec{r})$ for the two colliding nucleons, and the product of the two 
corresponding shadowing factors, according to Eq.~\ref{eq:5} and~\ref{eq:6}. By averaging the shadowing factors over all the \mbox{N-N} 
collisions in each \mbox{A-A} interaction, we get the centrality dependence of the shadowing factors. In Fig.~\ref{fig:3} we show, as a
function of $L$, the calculated shadowing factors for \mbox{In-In} and \mbox{Pb-Pb} collisions 
at $y$=0.5. For the SPS data at 158 GeV, this is the rapidity where the acceptance reaches its maximum. The symbols, connected by the 
continuous lines, have been calculated using Eq.~\ref{eq:5} for the local dependence of shadowing inside the nucleus, whereas the dashed 
line has been obtained using the parameterization of Eq.~\ref{eq:6}.
We notice that the two parametrizations give similar results, their difference not exceeding 2-3\%. In absence of other nuclear effects,
this result implies a $\sim$10\% anti-shadowing effect for central nucleus-nucleus collisions. When using EPS08, the effect increases 
up to $\sim$25\%.

\section{Comparison of cold nuclear matter effects in \lowercase{p}-A and \mbox{A-A} collisions}

We now want to discuss the extrapolation of cold nuclear matter effects measured in \mbox{p-A} to \mbox{A-A} collisions, taking into 
account shadowing effects together with final state absorption of the created $c\overline c$ pair. To do that, we model final state 
absorption effects using the simple law $\exp(-\rho\sigma^{abs}_{\rm J/\psi}L)$, which has been shown to reproduce, at first order, 
data on \jpsi\ production at fixed target energies. In this approach, the \jpsi\ production cross section per \mbox{N-N} collision 
in \mbox{p-A} is given by

\begin{equation}
\frac{1}{A}\sigma^{pA}_{\rm J/\psi}=\sigma^{NN}_{\rm J/\psi}\times S_{pA}^{\rm J/\psi}\times\exp(-\rho\sigma^{abs}_{\rm J/\psi}L)
\label{eq:7}
\end{equation}
 
We then fit  the \mbox{p-A} cross-section per \mbox{N-N} collision, obtained through Eq.~\ref{eq:7} with the simple 
$\exp(-\rho\sigma^{eff}_{\rm J/\psi}L)$ law, i.e. neglecting the existence of shadowing. This is the procedure usually followed at SPS 
energies to extract the so-called ``normal nuclear absorption''. Clearly this approach, when shadowing factors are not negligible, leads 
to two main consequences. On one hand the obtained  $\sigma^{eff}_{\rm J/\psi}$ values do not represent anymore the size of final-state 
absorption, but an effective quantity which, due to the presence of
shadowing~\cite{Lou09,Arl09}, may be quite different from $\sigma^{abs}_{\rm J/\psi}$. 
On the other hand, because shadowing effects do not necessarily have the same $L$-dependence of final state absorption, 
$\exp(-\rho\sigma^{eff}_{\rm J/\psi}L)$ may not
give anymore a reasonable fit of the \mbox{p-A} data. However, given the size of the shadowing corrections, their deviation from the
exponential behavior is difficult to observe in the existing data samples.

The \jpsi\ cross section per \mbox{N-N} collision in \mbox{A-A}, as a function of centrality, is then obtained using Eq.~\ref{eq:7}, 
replacing $S_{pA}^{\rm J/\psi}$  with $S_{AA}^{\rm J/\psi}$ and $1/A$ with $1/N_{\rm coll}$, where the shadowing factor and the
number of \mbox{N-N} collisions are calculated for the various centrality bins.
We can now compare the extrapolation of the exponential fit of the \mbox{p-A} calculation with what has been obtained for \mbox{A-A}
collisions. When doing that, we may expect significant deviations from
such an extrapolation. First of all, in \mbox{A-A} shadowing affects not only the target nucleus but also the projectile, leading to an
extra-effect that is clearly not present in the naive extrapolation of the exponential fit of \mbox{p-A}. Furthermore, these deviations
may heavily depend on rapidity, because when moving away from $y$=0 the $x$-region probed in the projectile becomes different from that of
the target.

\begin{figure}[htbp]
\centering
\resizebox{0.45\textwidth}{!}
{\includegraphics*[bb=0 0 530 530]{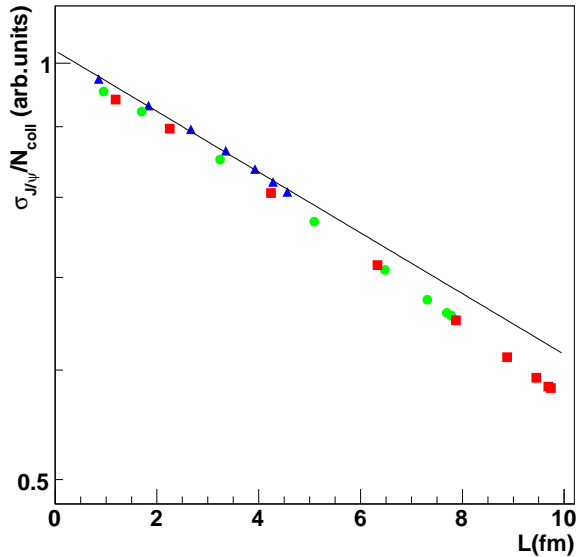}}
\caption{Comparison of \mbox{p-A} and \mbox {A-A} cross sections for \jpsi\ at 158 GeV, for $y$=0.5, including shadowing and final state
absorption, in arbitrary units. The \mbox{p-A} cross sections per \mbox{N-N} collisions are shown as triangles. The line represents an 
exponential fit to \mbox{p-A} data. \mbox{In-In} and \mbox{Pb-Pb} cross sections are shown as circles and squares, respectively.}
\label{fig:4}
\end{figure}

To illustrate the procedure, we show in Fig.~\ref{fig:4}, for 158 GeV energy and $y$=0.5, the expected behavior for \mbox{p-A}
collisions, including shadowing and having assumed $\sigma^{abs}_{\rm J/\psi}$=4 mb. We also show the result of an exponential fit to
\mbox{p-A} data, which gives $\sigma^{eff}_{\rm J/\psi}$=3.0 mb, a value smaller than $\sigma^{abs}_{\rm J/\psi}$, because of the presence of an
anti-shadowing effect. In the same plot we compare this exponential fit with the calculated cross section values per \mbox{N-N} 
collision for \mbox{In-In} and \mbox{Pb-Pb}, as a function of centrality. We clearly see that \mbox{A-A} cross sections deviate from the
extrapolation of \mbox{p-A} results.
In particular, at fixed $L$, the \jpsi\ cross section per \mbox{N-N} collision is systematically lower in \mbox{A-A} with respect to
\mbox{p-A}. However, we note that the relative behavior of \mbox{A-A} results with respect to \mbox{p-A}, at a certain $L$, cannot be easily 
deduced by a simple inspection of the \mbox{p-A} (and \mbox{A-p}) shadowing factors. In fact, from the geometry of the interaction, it can be
shown that the same $L$ for \mbox{p-A}~(\mbox{A-p}) and \mbox{A-A} corresponds to very different average nuclear densities probed in the 
collision and therefore to a different average strength of the shadowing effects. (To give a numerical example, the $L$ value
corresponding to \mbox{p-Pb} collisions is obtained with \mbox{Pb-Pb} collisions at $b=12$ fm. The average nuclear densities probed are
0.76$\rho_0$ and 0.48$\rho_0$ respectively, where $\rho_0$ is the core nuclear density).
Therefore, the \mbox{A-A} shadowing factors cannot be obtained as a simple product of \mbox{p-A} and \mbox{A-p} shadowing at the same $L$.

\begin{figure}[htbp]
\centering
\resizebox{0.45\textwidth}{!}
{\includegraphics*[bb=0 0 530 530]{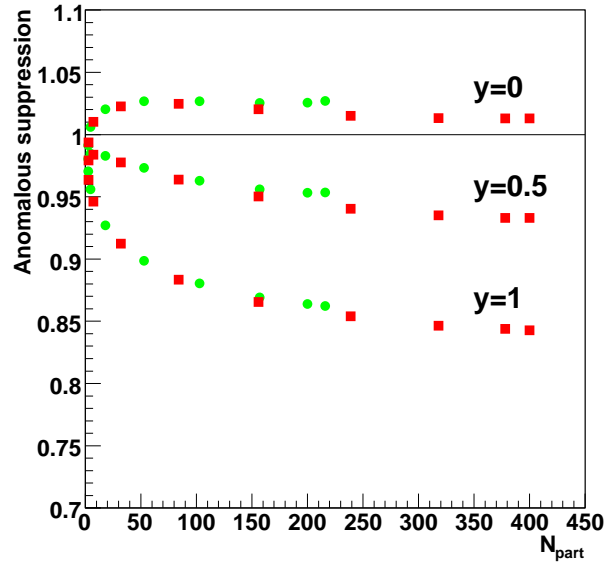}}
\caption{Ratio between the \mbox{A-A} cross section per \mbox{N-N} collision and the extrapolation of \mbox{p-A} results, for three
rapidity values, using the EKS98 parametrization. The results for \mbox{In-In} and \mbox{Pb-Pb} are shown as circles and squares, 
respectively.}
\label{fig:5}
\end{figure}

A deviation of \mbox{A-A} results from the \mbox{p-A} extrapolations was indeed found at SPS energies, and it was called 
``anomalous \jpsi\ suppression''~\cite{Ale05,Arn07}. Usually this effect was connected to hot nuclear matter effects, including the production of a
deconfined state~\cite{Gra02}. The result of Fig.~\ref{fig:4} shows that at least a fraction of this effect can be attributed to having
neglected the influence of shadowing in the determination of the ``nuclear absorption'' reference. In Fig.~\ref{fig:5} we present the ratio
between the \jpsi\ cross sections for \mbox{In-In} and \mbox{Pb-Pb} and the exponential extrapolation of \mbox{p-A} results, for three
rapidity values ($y$=0, 0.5 and 1), using EKS98 and Eq.~\ref{eq:5} for the local dependence of shadowing. 
This kinematical range corresponds to the region where \jpsi\ production has been 
studied by the NA50/NA60 experiments in nuclear collisions. The values of these ratios do not depend on the specific 
$\sigma^{abs}_{\rm J/\psi}$ value used in the calculation, because, at fixed $L$, the factors $\exp(-\rho\sigma^{abs}_{\rm J/\psi}L)$
cancel out in the ratio of the cross sections between \mbox{A-A} and \mbox{p-A}.

The values plotted in Fig.~\ref{fig:5} show that a simple extrapolation of \mbox{p-A} results, obtained through a fit of the 
$A$-dependence that parametrizes all nuclear effects under a single parameter $\sigma^{eff}_{\rm J/\psi}$, is not in agreement
with \mbox{A-A} \jpsi\ cross sections as a function of centrality.  This is basically because of the presence in \mbox{A-A} of shadowing 
effects in the projectile that are of course not present in \mbox{p-A}. Although the effect remains rather small at midrapidity, it
increases fast when moving away from $y$=0, reaching a discrepancy of $\sim$20\% between the extrapolation of \mbox{p-A} data and the
central \mbox{A-A} results for $y$=1. This effect must clearly be taken into account when looking for hot nuclear matter effects in
\mbox{A-A} data. In particular, the anomalous suppression values observed at the SPS must be rescaled by the values shown in
Fig.~\ref{fig:5} at the corresponding rapidities, resulting in a $\sim$10\% average reduction of this effect in the interval 0$<y<$1
where data were taken. When using the EPS08 parametrization, similar values to those of Fig.~\ref{fig:5} were obtained, with
relative discrepancies not larger than $\sim$2\%. 
However, it is well known that the uncertainties on the modification of gluon PDFs are 
quite large, and very recent analysis (EPS09) are now starting to systematically address this issue~\cite{Esk09}. The LO set of 
EPS09 gluon PDFs modification has an average value, in the $x$-region corresponding to SPS data, quite similar to the EKS98 one, with 
an error of the order of $\pm$15\%. By injecting such an uncertainty in our calculation, it turns out that the values shown in 
Fig.~\ref{fig:5} vary by about 5\%. Finally, other sets of nuclear modifications to gluon PDFs exist in the literature that exhibit either 
no (or little) anti-shadowing in our $x$-region (nDS/nDSg~\cite{deF04}) or an anti-shadowing very strongly increasing with $x$ 
(HKN~\cite{Hir07}). The use of such sets in our analysis gives almost no difference between
the extrapolation of \mbox{p-A} results and \mbox{A-A} for nDS/nDSg (ratios $\sim$1 in Fig.~\ref{fig:5}), or higher values, 
increasing with $y$, for \mbox{A-A} with respect to \mbox{p-A} for HKN (i.e. ratios larger than 1 in Fig.~\ref{fig:5}). 
However, it was pointed out~\cite{Esk09} that such analyses might be less constrained in the $x$-region under study, since
they do not make use of data from high-p$_{\rm T}$ $\pi$ production in \mbox{d-Au} collisions from the BNL Relativistic Heavy-Ion Collider 
(RHIC)~\cite{Adl07,Ada06}, which are
relevant for the determination of the large-$x$ region gluon contribution.

\section{Conclusions}

We have investigated in this article the role of shadowing for \jpsi\ production in \mbox{p-A} and \mbox{A-A} collisions at 
SPS energies. In particular, we have shown that an extrapolation of cold nuclear matter effects measured in \mbox{p-A} which does not take explicitly 
into account shadowing, cannot reproduce in a correct way such effects for \mbox{A-A}. In the frame of an LO Color Evaporation Model calculation, 
performed using the EKS98 and EPS08 parametrizations, we have shown that neglecting shadowing, the \mbox{p-A} extrapolation is biased by 
$\sim$10\% at $y$=0.5 for central \mbox{In-In} and \mbox{Pb-Pb} collisions. Such a bias must be taken into account in analysis that aim
at determining effects from hot nuclear matter in \jpsi\ production in \mbox{A-A} collisions at SPS energies 
(the so-called ``anomalous suppression''). In particular it may be quantitatively important for lighter systems such as \mbox{In-In}, where
the deviations from \mbox{p-A} extrapolations are relatively small.  



\begin{acknowledgments}
The authors wish to thank F.~Arleo, C.A.~Salgado and R.~Vogt for useful discussions on the topic covered in this article and for their comments
on the manuscript. 
We also gratefully acknowledge the help of D.~Berzano and F.~Prino for some of the computational aspects.
\end{acknowledgments}

	
\end{document}